\documentclass[twocolumn,showpacs,amsmath,amssymb,10pt]{revtex4}
\usepackage{graphicx,color}
\usepackage{amsmath}
\usepackage{amssymb}

\usepackage{amsfonts}
\usepackage{bm}

\newcommand{\ot}{{\,\otimes\,}}
\newcommand{{\Cd}}{{\mathbb{C}^d}}

\def\oper{{\mathchoice{\rm 1\mskip-4mu l}{\rm 1\mskip-4mu l}%
{\rm 1\mskip-4.5mu l}{\rm 1\mskip-5mu l}}}
\def\<{\langle}
\def\>{\rangle}
\newtheorem{Theorem}{Theorem}

\newtheorem{remark}{Remark}
\newtheorem{Example}{Example}
\newtheorem{Proposition}{Proposition}
\newcommand{\REV}[1]{{\color{red}[[SAVERIO: #1]]}}
\newcommand{\BLUE}[1]{{\color{blue}[[DAREK: #1]]}}

\begin{document}
\title{\textbf{From Markovian semigroup to non-Markovian quantum evolution}}
\author{Dariusz Chru\'sci\'nski and Andrzej Kossakowski}
\affiliation{Institute of Physics, Nicolaus Copernicus University \\
Grudzi\c{a}dzka 5/7, 87--100 Toru\'n, Poland}

\begin{abstract}
We  provided a class of legitimate memory kernels leading to
completely positive trace preserving dynamical maps. Our
construction is based on a simple normalization procedure.
Interestingly, when applied to the celebrated Wigner-Weisskopf
theory it gives the standard Markovian evolution governed by the
local master equation.

\end{abstract}


\maketitle

\section{Introduction}

The dynamics of open quantum systems attracts nowadays increasing
attention \cite{Breuer,Weiss,Alicki}.  It is  relevant not only for
the better understanding of quantum theory but it is fundamental in
various modern applications of quantum mechanics. Since the
system-environment interaction causes dissipation, decay and
decoherence it is clear that dynamic of open systems is fundamental
in modern quantum technologies, such as quantum communication,
cryptography and computation \cite{QIT}.

The usual approach to the dynamics of an open quantum system
consists in applying the Born-Markov approximation, that leads to
the following master equation
\begin{equation}\label{M}
 \frac{d}{dt} \rho_t = L \rho_t\ ,     \ \ \ \rho_0=\rho\ ,
\end{equation}
where $L$ denotes the corresponding  Markovian generator (see
Section \ref{Markov} for all details). However, it turns out that
description of many complex systems requires more sophisticated
analysis which take into account non-Markovian memory effects
\cite{Breuer}. A popular non-Markovian generalization of (\ref{M})
is the following nonlocal equation
\begin{equation}\label{NM}
    \dot{\rho}_t = \int_0^t \mathcal{K}_{t-\tau}\, \rho_\tau \, d\tau\ ,
\end{equation}
in which quantum memory effects are taken into account through the
introduction of the memory kernel $\mathcal{K}_\tau$. It is clear
from (\ref{NM}) that the rate of change of the state $\rho_t$ at
time $t$ depends on its history (starting at $t=0$). The Markovian
master equation (\ref{M}) is reobtained when $\mathcal{K}_\tau =
2\delta(\tau)L$. The time dependent kernel $\mathcal{K}_t$ is
usually referred to as the generator of the non-Markovian master
equation.

Non-Markovian systems appear in many branches of physics, such as
quantum optics \cite{Breuer,Gardiner}, solid state physics
\cite{solid}, quantum chemistry \cite{Plenio-K}, and quantum
information processing \cite{Aharonov}. Since non-Markovian dynamics
modifies exponential decay of quantum coherence it turns out that
when applied to composite systems it may protect quantum
entanglement for longer time than standard Markovian evolution
\cite{Saverio}. In particular it may protect the system against the
sudden death of entanglement \cite{DEATH}. Non-Markovian dynamics
was recently studied in
\cite{Wilkie,Budini,B-2004,Wodkiewicz,Lidar,Maniscalco1,Maniscalco2,Maniscalco-09,KR,KR-last,B,Francesco,local}.
Interestingly, several measures of non-Markovianity were proposed
during last year \cite{Wolf,Breuer-09,Plenio,Chiny-09}.


One of the fundamental problems in the theory of non-Markovian
master equations is to find those conditions on $\mathcal{K}_t$ that
ensure that the time evolution $\Lambda_t$ defined by
\begin{equation}\label{}
    \rho \ \longrightarrow\ \rho_t = \Lambda_t \rho
\end{equation}
is completely positive and trace preserving
\cite{Wilkie,Budini,B-2004,KR,KR-last,B}. This problem is very
involved and contrary to the Markovian case the full
characterization of the corresponding properties of memory kernel is
still unknown.

In the present paper we provide a class of memory kernels giving
rise to legitimate quantum dynamics. Our construction is based on a
simple ide of normalization: starting from a family of completely
positive maps satisfying a certain additional condition one is able
to `normalize' it in order to obtain legitimate, i.e. trace
preserving, quantum dynamics. As a result on obtains a class of
legitimate memory kernels.

The paper is organized as follows. In Section \ref{Bernstein} we
provide the hierarchy of necessary conditions for the memory kernel
which guarantee the legitimate quantum dynamics. Section
\ref{Markov} discusses the structure of Markovian semigroup and
introduces basic idea of normalization. It turns out that Markovian
semigroup appears as a normalized Wigner-Weisskopf theory. Section
\ref{MAIN} provides the main body of the paper. We show that there
is a natural way to construct a legitimate memory kernel {\em via }
an appropriate normalization procedure. Then in Section \ref{CLASS}
as a byproduct we provide the construction of legitimate memory
kernels in classical stochastic non-Markovian dynamics. In Section
\ref{WW} we analyze the reduction of the Schr\"odinger dynamics in
the Hilbert space `system + reservoir'. It turn out that reduced
dynamics may be normalized to legitimate non-Markovian dynamics in
the space of density operators. Final conclusions are collected in
the last Section.

\section{Memory kernels and quantum Bernstein theorem}
\label{Bernstein}

A solution $\Lambda_t$ to the non-Markovian master equation is trace
preserving iff ${\rm Tr}( \mathcal{K}_t \rho) = 0$ for any density
operator $\rho$. Equivalently, this condition may be rewritten in
terms of the dual of $\mathcal{K}_t$ as follows
\begin{equation}\label{TR}
\mathcal{K}_t^\# \mathbb{I} = 0 \ ,
\end{equation}
where $\mathbb{I}$ denotes an identity operator (recall that if $K$
is a linear map $K : \mathcal{B}(\mathcal{H}) \longrightarrow
\mathcal{B}(\mathcal{H})$, then its dual $K^\#$ is defined by ${\rm
Tr}(K^\# a \cdot b ) = {\rm Tr}(a\cdot K b)$ for any $a,b
\in\mathcal{B}(\mathcal{H})$).

Let us turn to more difficult part, i.e. complete positivity of
$\Lambda_t$.  Let us observe that taking the Laplace transform of
(\ref{NM}) one obtains
\begin{equation}\label{Ls}
    \widetilde{\Lambda}_s = \frac{1}{s - \widetilde{\mathcal{K}}_s}
    \ ,
\end{equation}
where
\begin{equation}\label{Ls1}
    \widetilde{\Lambda}_s = \int_0^\infty e^{-st} \Lambda_t dt\ ,
\end{equation}
and $s \in \mathbb{C}$. Now, if $\Lambda_t$ is completely positive
then for $s > 0$ its Laplace transform $ \widetilde{\Lambda}_s$ is
completely positive as well.  Observe that
\begin{equation}\label{Ls2}
    (-1)^k \frac{d^k}{ds^k} \, \widetilde{\Lambda}_s = \int_0^\infty e^{-st} t^k \Lambda_t dt\
    .
\end{equation}
Now, the r.h.s of (\ref{Ls2}) is completely positive being a convex
combination of completely positive maps $\Lambda_t$ (for $s > 0$).
Hence, using (\ref{Ls}) one finds that for any positive integer $k$
and $s >0 $
\begin{equation}\label{Ls4}
 (-1)^k \frac{d^k}{ds^k} \, \frac{1}{s - \widetilde{\mathcal{K}}_s}\
 \ \ \ \ {\rm is\ completely\ positive}\ .
\end{equation}
 This provides a series of
necessary conditions for $\mathcal{K}_t$. Note, that these
conditions may be considered as a quantum version of Bernstein
theorem \cite{Bernstein1,Bernstein2}. Recall, that a function $f :
[\,0,\infty) \rightarrow \mathbb{R}$ is completely monotone if
\begin{equation}\label{}
    (-1)^k \frac{d^k}{dt^k} \, f(t) \geq 0 \ ,
\end{equation}
for all $t\in [0,\infty)$ and $k=0,1,2,\ldots$. Then Bernstein
theorem states that $f$ is completely monotone if and only if $f$ is
a Laplace transform of the non-negative function, that is,
\begin{equation}\label{}
    f(s) = \int_0^\infty e^{-st} g(t) dt\ ,
\end{equation}
with $g(t) \geq 0$ and positive $s$. Unfortunately, apart from
mathematical elegance the above infinite hierarchy of necessary
conditions are very hard to use in practice. In particular if
$\mathcal{K}_t$ is non-commutative family, that is,
$[\mathcal{K}_t,\mathcal{K}_\tau]\neq 0$ for $t \neq \tau$, then
even simple differentiation in (\ref{Ls4}) is by no means trivial.
Therefore, one needs other tools to analyze properties of
$\mathcal{K}_t$ which guarantee that $\Lambda_t$ is a legitimate
quantum evolution with memory.

\section{The structure of Markovian dynamics}  \label{Markov}

To present our ideas we start with Markovian semigroup. As is well
known \cite{GKS,Lindblad} the most general structure of the
Markovian master equation is given by
\begin{equation}\label{M1}
 \frac{d}{dt} \rho_t = L \rho_t\ ,     \ \ \ \rho_0=\rho\ ,
\end{equation}
where the Markovian generator  $L$ is given by
\begin{equation}\label{}
    L \, \rho = -i[H,\rho] + \frac 12 \sum_\alpha \left( [V_\alpha,\rho
    V_\alpha^\dagger] + [V_\alpha\rho,V_\alpha^\dagger] \right) \ .
\end{equation}
In the above formula $H$ represents system Hamiltonian and
$\{V_\alpha \}$ is the collection of arbitrary operators encoding
the interaction between system and the environment. Let us observe
that $L$ may be rewritten in the following form
\begin{equation}\label{LBZ}
    L = B - Z \ ,
\end{equation}
where $B$ is a completely positive map defined by the following
Kraus representation
\begin{equation}\label{}
    B\, \rho = \sum_\alpha V_\alpha\rho V_\alpha^\dagger\ ,
\end{equation}
and the super-operator $Z$ reads as follows
\begin{equation}\label{}
    Z\, \rho = -i(C \rho  - \rho\, C^\dagger)\ ,
\end{equation}
with
\begin{equation}\label{C}
    C = H - \frac i2 B^\# \mathbb{I}  \ .
\end{equation}
Note, that $B^\#$ denotes the dual map
\begin{equation}\label{}
B^\# a = \sum_\alpha V_\alpha^\dagger a V_\alpha\ ,
\end{equation}
and hence $B^\# \mathbb{I} = \sum_\alpha V_\alpha^\dagger V_\alpha$
satisfies $B^\# \mathbb{I} \geq 0$. Note, that by construction
\begin{equation}\label{BZ-I}
    B^\# \mathbb{I} =  Z^\# \mathbb{I}\ ,
\end{equation}
which implies that $L^\# \mathbb{I} = 0$, and hence the dynamics
$\Lambda_t$ preserves the trace.

Now comes the natural question: which term in (\ref{LBZ}) is more
fundamental $B$ or $Z$? Clearly, knowing completely positive $B$ the
$Z$ part is up to the Hamiltonian part uniquely defined. On the
other hand if $B=0$, then  $Z\rho = - i[H,\rho]$ reduces to the
purely Hamiltonian part. Hence, in our opinion, `$Z$ part' plays a
primary role replacing Hamiltonian $H$ by non-Hermitian operator $C$
\begin{equation}\label{CHX}
    H \ \rightarrow \ C = H - \frac i2 X\ ,
\end{equation}
with $X \geq 0$, that is, one introduces non-Hermitian Hamiltonian
$C$  giving rise to non-unitary dynamics. This approach is the heart
of celebrated Wigner-Weisskopf theory \cite{WW}. Actually, in the
standard Wigner-Weisskopf theory $C$ is normal ($CC^\dagger =
C^\dagger C$) which means that $[H,X]=0$. In this case there is an
orthonormal basis $|k\>$ in $\mathcal{H}$ such that
\begin{equation}\label{}
H|k\> = E_k |k\>  \ , \ \ \ \  X|k\> = \Gamma_k |k\> \ ,
\end{equation}
with $\Gamma_k \geq 0$, for $k=1,2,\ldots,$dim$\mathcal{H}$.
Therefore, if $|\psi_0\> = \sum_k c_k |k\>$, then
\begin{equation}\label{}
|\psi_t\> = e^{-iC t} |\psi_0\> =  \sum_k c_k(t) |k\>\ ,
\end{equation}
 with
\begin{equation}\label{}
    c_k(t) = e^{-(iE_k + \frac 12 \Gamma_k)t}\, c_k\ ,
\end{equation}
and hence one recovers celebrated exponential decay. We stress,
however, that $C$ needs not be normal. The only essential thing is
that $X$ in (\ref{CHX}) is positive semidefinite. Actually, there is
big activity in the field of non-Hermitian Hamiltonians displaying
real spectra (see e.g. recent review by Bender \cite{Bender}).

Now, let $N_t$ be solution to
\begin{equation}\label{N1}
    \dot{N}_t = - Z N_t\ , \ \ \ N_0=\oper\ .
\end{equation}
One easily finds
\begin{equation}\label{NCC}
    N_t \rho = e^{-iCt} \rho \,e^{iC^\dagger t} \ .
\end{equation}
It is evident that $N_t$ is completely positive. Note however that
it does not preserve the trace (unless $X=0$). One finds
\begin{equation}\label{}
    N^\#_t \mathbb{I} = e^{-iCt} e^{iC^\dagger t}\ ,
\end{equation}
and if $C$ is normal it simplifies to $
 N^\#_t \mathbb{I} = e^{- Xt}$.
Interestingly,
\begin{equation}\label{}
    \dot{N}^\#_t \mathbb{I} = - e^{-iCt}Xe^{iC^\dagger t}\ ,
\end{equation}
and hence
\begin{equation}\label{}
-    \dot{N}^\#_t \mathbb{I}  \geq 0\ .
\end{equation}
This condition would play a crucial role in our analysis of
non-Markovian evolution. Here, we point out that it is trivially
satisfied in the Markovian case.

 It is, therefore, clear the `$B$ term' is needed just to
normalize evolution. Let us observe the the choice of completely
positive $B$ is highly non unique. The only condition for $B$ is
$B^\# \mathbb{I} = X$.

Finally, the Laplace transforms of (\ref{M1}) and (\ref{N1}) give
\begin{equation}\label{Lambda-N-M}
    \widetilde{\Lambda}_s = \frac{1}{s + Z - B }\ , \ \ \ \widetilde{N}_s = \frac{1}{s + Z}\ ,
\end{equation}
and hence on obtains the following relation
\begin{equation}\label{}
\widetilde{\Lambda}_s = \widetilde{N}_s + \widetilde{N}_s B
\widetilde{\Lambda}_s\ .
\end{equation}
Iterating this equation yields the following perturbation series
\begin{equation}\label{Lambda-series-M}
\widetilde{\Lambda}_s = \widetilde{N}_s + \widetilde{N}_s B
\widetilde{N}_s + \widetilde{N}_s B \widetilde{N}_s B
\widetilde{N}_s + \ldots\ .
\end{equation}
Now, since both $\widetilde{N}_s$ and $B$ are completely positive it
is clear from (\ref{Lambda-series-M}) that $\widetilde{\Lambda}_s$
is completely positive. Going back to the time-domain it finally
shows that $\Lambda_t$ defines legitimate quantum evolution.

Note, that analyzing the series (\ref{Lambda-series-M}) one is
tempted to relax the condition of complete positivity for $B$. It is
evident that it is sufficient that $B\widetilde{N}_t$ is completely
positive. Note, however, that due to (\ref{NCC}) the map $N_t$ is
invertible and the inverse
\begin{equation}\label{}
    N^{-1}_t \rho = e^{iCt} \rho \,e^{-iC^\dagger t} \ ,
\end{equation}
is completely positive as well. Hence $(BN_t)N_t^{-1}$ is again
completely positive. But  $(BN_t)N_t^{-1} = B$ which shows that
complete positivity of $B$ cannot be relaxed.

\section{A class of legitimate memory kernels}  \label{MAIN}

In this section which provides the main body of the paper we
generalize `normalizing procedure' from semigroups to non-Markovian
dynamics. We shall consider a class of non-Markovian Master
Equations
\begin{equation}\label{Lambda-t}
    \frac{d}{dt}\, \Lambda_t = \int_0^t \mathcal{K}_{t-\tau} \Lambda_\tau\, d\tau\ ,\
    \ \ \ \rho_0=\rho\ ,
\end{equation}
assuming that the memory kernel $\mathcal{K}_t$ -- in analogy to
(\ref{LBZ}) -- can be represented in the following form
\begin{equation}\label{K-1}
    \mathcal{K}_t = B_t - Z_t\ .
\end{equation}
Unfortunately, we do not know how to chose $B_t$ and $Z_t$ in order
to generate legitimate dynamical map $\Lambda_t$. Note that to
satisfy (\ref{TR}), one has the following constraint
\begin{equation}\label{L-2}
    B^\#_t \mathbb{I} = Z_t^\# \mathbb{I}\ .
\end{equation}
The most difficult part is to guarantee that $\Lambda_t$ is
completely positive for all $t \geq 0$.

To find legitimate $B_t$ and $Z_t$ we propose the following
procedure: let us introduce an arbitrary  map $N_t$ which is
completely positive and satisfies initial condition $N_0 = \oper$.
Note, that $N_t$ may be represented as follows
\begin{equation}\label{NF}
    N_t = \oper - \int_0^t F_\tau d\tau\ ,
\end{equation}
where $F_t = -\dot{N}_t$.  Assuming that $N_t$ is differentiable it
always satisfies the following non-local equation
\begin{equation}\label{N-t}
    \frac{d}{dt} N_t = - \int_0^t Z_{t-\tau} N_\tau \, d\tau\ , \ \
    \ N_0 = \oper\ ,
\end{equation}
where the corresponding generator $Z_t$ is defined in terms of its
Laplace transform
\begin{equation}\label{Z-N}
    \widetilde{Z}_s = \frac{\oper - s
    \widetilde{N}_s}{\widetilde{N}_s}\ .
\end{equation}
 Hence, any family of completely positive
maps $N_t$ provides $Z_t$ in (\ref{K-1}). To provide $B_t$
satisfying condition (\ref{L-2}) let us observe that equation
(\ref{Lambda-t}) implies
\begin{equation}\label{Lambda-N-nM}
    \widetilde{\Lambda}_s = \frac{1}{s + \widetilde{Z}_s - \widetilde{B}_s }\
    .
\end{equation}
Moreover, one gets from (\ref{Z-N})
\begin{equation}\label{N-Z}
\widetilde{N}_s = \frac{1}{s + \widetilde{Z}_s}\ .
\end{equation}
Hence
\begin{equation}\label{}
\widetilde{\Lambda}_s = \widetilde{N}_s + \widetilde{N}_s
\widetilde{B}_s \widetilde{\Lambda}_s\ .
\end{equation}
Iterating this equation yields the following perturbation series
\begin{equation}\label{Lambda-series-nM}
\widetilde{\Lambda}_s = \widetilde{N}_s + \widetilde{N}_s
\widetilde{B}_s \widetilde{N}_s + \widetilde{N}_s \widetilde{B}_s
\widetilde{N}_s \widetilde{B}_s \widetilde{N}_s + \ldots\ .
\end{equation}
This equation generalizes the Markovian formula
(\ref{Lambda-series-M}). It is therefore clear that if the map
$\widetilde{B}_s \widetilde{N}_s$, or equivalently in the time
domain $\int_0^t B_{t-\tau} N_\tau d\tau$ is completely positive,
then due to (\ref{Lambda-series-nM}) the corresponding dynamical map
$\Lambda_t$ is completely positive as well. Let us assume that the
family of completely positive maps $N_t$ is invertible (clearly,
$N^{-1}_t$ needs not be completely positive). Now, let us define
$B_t$ is terms of its Laplace transform
\begin{equation}\label{}
    \widetilde{B}_s = \widetilde{Q}_s \widetilde{N}^{-1}_s \ ,
\end{equation}
where $\widetilde{Q}_s$ denotes the Laplace transform of the
completely positive map $Q_t$. Equivalently, one has the following
prescription for the Laplace transform of the memory kernel
\begin{equation}\label{}
    \widetilde{\mathcal{K}}_s = \widetilde{Q}_s \widetilde{N}^{-1}_s - \widetilde{Z}_s\
    .
\end{equation}
Now, $\widetilde{B}_s \widetilde{N}_s = \widetilde{Q}_s$ is by
construction completely positive and hence due to
(\ref{Lambda-series-nM}) the above memory kernel generates
completely positive dynamics. One obtains
\begin{equation}\label{KK}
    \widetilde{\mathcal{K}}_s = [\widetilde{Q}_s - (\oper - s \widetilde{N}_s)]
    \widetilde{N}^{-1}_s \ ,
\end{equation}
and hence
\begin{equation}\label{}
\widetilde{\mathcal{K}}_s^\# \mathbb{I} = \widetilde{N}^{-1
\#}_s[\widetilde{Q}^\#_s - (\oper - s
\widetilde{N}^\#_s)]\mathbb{I}\ .
\end{equation}
Now, if
\begin{equation}\label{QN-1}
[\widetilde{Q}^\#_s - (\oper - s \widetilde{N}^\#_s)]\mathbb{I} = 0
\ ,
\end{equation}
then $\widetilde{K}_t$ defines legitimate  memory kernel. Recalling,
that $\oper - s \widetilde{N}_s$ corresponds to the Laplace
transform of $d{N}_t/dt$ one may rewrite (\ref{QN-1}) as follows
\begin{equation}\label{QN-2}
    Q^\#_t \mathbb{I} + \dot{N}^\#_t \mathbb{I} = 0\ ,
\end{equation}
or, using (\ref{NF})
\begin{equation}\label{QN-3}
    Q^\#_t \mathbb{I} = {F}^\#_t \mathbb{I} \ .
\end{equation}
Now,it is clear that if $F^\#_t \mathbb{I} \geq 0$ for all $t\geq0$,
then one can always find completely positive $Q_t$ satisfying
(\ref{QN-3}).

 Summarizing, we proved the following

\begin{Theorem}
Let $N_t$ be an arbitrary (differentiable) family of completely
positive maps satisfying (\ref{NF}) such that $F^\#_t \mathbb{I}
\geq 0$. Then, there exists a family of completely positive maps
$Q_t$ satisfying (\ref{QN-3}) and the formula (\ref{KK}) defines the
Laplace transform of the legitimate generator of non-Markovian
dynamical map.
\end{Theorem}

It should be clear that the construction of $Q_t$ is highly non
unique. If $F^\#_t \mathbb{I} \geq 0$, then there is a
time-dependent family of operators $M_\alpha(t)$ such that
\begin{equation}\label{}
F^\#_t \mathbb{I}  = \sum_\alpha M^\dagger_\alpha(t) M_\alpha(t) \ ,
\end{equation}
and hence one can define $Q_t$ via the following Kraus
representation
\begin{equation}\label{}
    Q_t \rho = \sum_\alpha M_\alpha(t) \rho M^\dagger_\alpha(t) \ .
\end{equation}
Again, the choice of $M_\alpha(t)$ is highly non unique.  Note, that
the simplest way to satisfy (\ref{QN-3}) is to take
\begin{equation}\label{QBF}
    Q_t = B F_t \ ,
\end{equation}
where $B$ denotes a quantum channel (completely positive trace
preserving map). Indeed, one has $Q^\#_t \mathbb{I} = F_t^\# (B^\#
\mathbb{I}) = F_t^\# \mathbb{I}$, due to $B^\# \mathbb{I} =
\mathbb{I}$. In this case one obtains the following form of the
memory kernel
\begin{equation}\label{KBZ}
    \mathcal{K}_t = (B-\oper)Z_t\ .
\end{equation}

\begin{Example} {\em
Let us observe that $Q_t$ defined via (\ref{QBF}) is completely
positive whenever $F_t$ is completely positive.  Note, however, that
due to (\ref{NF}) the map $F_t$ can not be completely positive
unless $F_t = f(t) \oper$ for some non-negative function $f$.
Indeed, the corresponding Choi matrices for $N_t$, $F_t$ and
identity map $\oper$ have to be positive. Now, the Choi matrix for
the identity map defines rank-1 projector $P^+$ (the maximally
entangled state in $\mathcal{H} \ot \mathcal{H}$) and therefore it
is clear that we can not subtract from $P^+$ any positive operator
unless it is proportional to $P^+$ itself (otherwise the Choi matrix
for $N_t$ would no longer be positive). Hence
\begin{equation}\label{}
    N_t = \left( 1 - \int_0^t f(\tau)\, d\tau \right) \oper\ ,
\end{equation}
where $f(t) \geq 0$, and to guarantee complete positivity of $N_t$
one has
\begin{equation}\label{}
 \int_0^\infty f(t)\, dt \leq 1\ .
\end{equation}
Finally, using (\ref{Z-N}) one finds the following formula
\begin{equation}\label{}
    Z_t = \kappa(t)\, \oper\ ,
\end{equation}
where the function $\kappa(t)$ is defined in terms of its Laplace
transform
\begin{equation}\label{}
    \widetilde{\kappa}(s) = \frac{s \widetilde{f}(s)}{1-
    \widetilde{f}(s)} \ .
\end{equation}
To find $B_t$ we define its Laplace transform by $\widetilde{B}_s =
Q_s \widetilde{N}^{-1}_s$, where $Q_t$ are arbitrary completely
positive maps. One has
\begin{equation}\label{}
    \widetilde{N}^{-1}_s = \frac{s}{1-\widetilde{f}(s)} \, \oper
     ,
\end{equation}
and hence
\begin{equation}\label{}
    \widetilde{K}_s = \frac{s}{1-\widetilde{f}(s)} \Big[\, \widetilde{Q}_s -
 \widetilde{f}(s) \oper\, \Big] \ .
\end{equation}
Note, that to satisfy (\ref{TR}) one has
\begin{equation}\label{}
    \widetilde{Q}_s^\# \mathbb{I} = \widetilde{f}(s) \mathbb{I}\ .
\end{equation}
In particular if $Q_t = BF_t$, then $\widetilde{Q}_s =
\widetilde{f}(s)B$ and the memory kernel has the following form
\begin{equation}\label{}
    \mathcal{K}_t = \kappa(t) (B - \oper)\ ,
\end{equation}
with $B$ being an arbitrary quantum channel.

}
\end{Example}

\begin{Example} {\em
Consider now the following class of completely positive maps $N_t$
in $\mathcal{B}(\mathcal{H})$ with $\mathcal{H}= \mathbb{C}^d$
\begin{equation}\label{N-nij}
    N_t\, \rho = \sum_{i,j=1}^d n_{ij}(t)\, |i\>\<i|\,\rho\, |j\>\<j|\
    ,
\end{equation}
where the matrix $[n_{ij}(t)] \geq 0 $ for $t\geq 0$, and
$n_{ij}(0)=1$ which guaranties that $N_0 = \oper$. Equation
(\ref{NF}) implies the following formula for the corresponding map
$F_t$
\begin{equation}\label{}
    F_t\, \rho = \sum_{i,j=1}^d f_{ij}(t)\, |i\>\<i|\,\rho\, |j\>\<j|\ ,
\end{equation}
where
\begin{equation}\label{f-ij}
    f_{ij}(t) = - \frac{d\, n_{ij}(t)}{dt}\ .
\end{equation}
Let $Q_t$ be a family of complete positive maps defined by the
corresponding Kraus representation
\begin{equation}\label{}
Q_t\, \rho = \sum_{i,j,k,l=1}^d q_{ij;kl}(t)\, |i\>\<j|\,\rho
|l\>\<k| \ .
\end{equation}
It is clear that $Q_t$ is completely positive iff
\begin{equation}\label{}
\sum_{i,j,k,l=1}^d q_{ij;kl}(t)\, x_{ij}\, \overline{x_{kl}} \geq 0
\ ,
\end{equation}
for any $d \times d$ complex matrix $[x_{ij}]$. One has
\begin{equation}\label{}
F^\#_t \mathbb{I} = \sum_k f_{kk}(t) |k\>\<k|\ ,
\end{equation}
and hence  $F^\#_t \mathbb{I} \geq 0$ if $f_{kk}(t) = -
\dot{n}_{kk}(t) \geq 0$. One finds for $Q^\#_t \mathbb{I}$
\begin{equation}\label{}
Q^\#_t \mathbb{I} = \sum_{i,j,k} q_{ij;kj}(t) \, |k\>\<i|\ .
\end{equation}
Now, to satisfy (\ref{QN-3}) one has
\begin{equation}\label{}
    q_{ij;kl}(t) = \delta_{ik} \, c^{(k)}_{jl}(t)\ ,
\end{equation}
where the time-dependent $d \times d$ complex matrices $c^{(k)}(t)$
are positive semi-definite, i.e. they define unnormalized density
operators. Finally, to satisfy (\ref{QN-3}) one has the following
conditions
\begin{equation}\label{}
    {\rm Tr}\, c^{(k)}(t) = f_{kk}(t)\ ,
\end{equation}
for $k=1,2,\ldots,d$.

One may ask how to construct (\ref{N-nij}) in order to satisfy
$\dot{n}_{kk}(t) \leq 0$. Here we present the following
construction: let  $X_1,\ldots,X_d$ be a set of arbitrary linear
operators from $\mathcal{B}(\mathcal{H})$. Define
\begin{equation}\label{}
    n_{ij}(t) = {\rm Tr} \left( \omega e^{tX_i^\dagger} e^{t X_j}
    \right) \ ,
\end{equation}
where $\omega$ is a fixed density operator. By construction
$[n_{ij}(t)]\geq 0$. Moreover, $n_{ij}(0) = {\rm Tr}\, \omega = 1$.
One obtains
\begin{equation}\label{}
    f_{ij}(t) = - \dot{n}_{ij}(t) = - {\rm Tr} \left(  \omega\, e^{tX_i^\dagger}(X_i^\dagger + X_j) e^{t X_j}
    \right) \ ,
\end{equation}
and hence
\begin{equation}\label{}
    f_{ii}(t) = - {\rm Tr} \left( \omega_i\,(X_i^\dagger + X_i)
    \right) \ ,
\end{equation}
where
\begin{equation}\label{}
    \omega_i = e^{t X_i}\, \omega\, e^{tX_i^\dagger}\ .
\end{equation}
Now, if each $X_i$ is dissipative, i.e. $X_i + X_i^\dagger \leq 0$,
then one gets $f_{ii}(t) \geq 0$.

}
\end{Example}

\begin{remark} {\em
Let us observe hat there is another way to normalize the family of
completely positive maps $N_t$. Suppose that $X_t := N_t^\#
\mathbb{I}
> 0 \,$, for all $t \geq 0$ and define
\begin{equation}\label{N-II}
    M^\#_t a = X_t^{-1/2} ( N_t^\# a ) X_t^{-1/2}\ .
\end{equation}
One has $ M^\#_t \mathbb{I} = \mathbb{I}$, and hence $M_t$ defines a
legitimate dynamical map. Note, however, that we are not able to
write down the corresponding equation for $M_t$. Moreover, the above
normalization is again highly non unique. If $U_t$ is an arbitrary
family of unitary operators, then $a \rightarrow U_t (M^\#_t a )
U^\dagger_t$ does preserve $\mathbb{I}$. }
\end{remark}

\section{Classical non-Markovian dynamics} \label{CLASS}

As a byproduct of our general approach one obtains a coherent
description of classical stochastic dynamics. A mixed state of a
$d$-state classical system is described by a stochastic $d$-vector
$(p_1,\ldots,p_d)$. Any such vector may be encoded into diagonal
density operator $\rho_{kl} = p_k \delta_{kl}$. We call a linear map
in $\mathcal{B}(\mathcal{H})$ to be classical if it maps diagonal
matrices into diagonal matrices (in a fixed basis in $\mathcal{H}$).

Consider now a classical completely positive maps
\begin{equation}\label{}
    N_t \rho = \sum_{i,j=1}^d n_{i}(t) |i\>\<i| \, \rho\, |i\>\<i|
    \ ,
\end{equation}
which is nothing but the `diagonal part' of (\ref{N-nij}). Moreover,
one assumes that  $n_i(t) \geq 0$ and $n_i(0)=1$. Applying to
probability vector the action of $N_t$ is very simple: it maps $p_k$
into $n_k(t) p_k$. Clearly, $N_t$ is not normalized: $N^\#_t
\mathbb{I} = \sum_k n_k(t) |k\>\<k|$. Moreover, it defines $Z_t$
\begin{equation}\label{}
 Z_t \rho = \sum_{i,j=1}^d z_{i}(t) |i\>\<i| \, \rho\, |i\>\<i|
    \ ,
\end{equation}
where $z_{i}(t)$ are defined in terms of the Laplace transform
\begin{equation}\label{}
    \widetilde{z}_{i}(s) =
    \frac{1-s\widetilde{n}_{i}(s)}{\widetilde{n}_i(s)}\ .
\end{equation}
Now, let $Q_t$ be another family of classical completely positive
maps
\begin{equation}\label{}
    Q_t \rho = \sum_{i,j=1}^d q_{ij}(t) |i\>\<j| \, \rho\, |j\>\<i|\
    ,
\end{equation}
where the time-dependent coefficients satisfy $q_{ij}(t) \geq 0$ for
$t \geq 0$. Normalization condition (\ref{QN-2}) implies
\begin{equation}\label{qn}
    \sum_{i=1}^d q_{ij}(t) + \dot{n}_j(t) = 0\ ,
\end{equation}
for $j=1,\ldots,d$. Let us introduce $B_t$ according to
$\widetilde{B}_s = \widetilde{Q}_s \widetilde{N}^{-1}_s$. One has
\begin{equation}\label{}
    B_t \rho = \sum_{i,j=1}^d b_{ij}(t) |i\>\<j| \, \rho\, |j\>\<i|\
    ,
\end{equation}
where $b_{ij}(t)$ are defined in terms of the Laplace transform
\begin{equation}\label{}
    \widetilde{b}_{ij}(s) =
    \frac{\widetilde{q}_{ij}(s)}{\widetilde{n}_j(s)}\ .
\end{equation}
Finally, one arrives to the following formula for the memory kernel
\begin{equation}\label{}
    \mathcal{K}_t \rho = \sum_{i,j=1}^d k_{ij}(t) |i\>\<j| \, \rho\, |j\>\<i|\
    ,
\end{equation}
where $k_{ij}(t)$ are defined as follows
\begin{equation}\label{}
    k_{ij}(t) = b_{ij}(t) - \delta_{ij} z_j(t) \ .
\end{equation}
Observe, that (\ref{qn}) implies $\sum_{i=1}^d k_{ij}(t)=0$ and
hence
\begin{equation}\label{}
\sum_{i=1}^d b_{ij}(t) = z_j(t)\ .
\end{equation}
When translated into the stochastic vector our approach gives rise
to the following classical non-Markovian master equation
\begin{equation}\label{NM-CLASS}
    \dot{p}_i(t) = \sum_{j=1}^d \int_0^t d\tau \Big[ b_{ij}(t-\tau)
    p _j(\tau) - b_{ji}(t-\tau) p_i(\tau) \Big] \ .
\end{equation}
Let us consider the special case corresponding to (\ref{QBF}). One
introduces $B$ by
\begin{equation}\label{}
 B \rho = \sum_{i,j=1}^d \pi_{ij} |i\>\<j| \, \rho\, |j\>\<i|\
    ,
\end{equation}
where $[\pi_{ij}]$ is a stochastic matrix, i.e. $\pi_{ij} \geq 0$,
and $\sum_i \pi_{ij} = 1$. One finds
\begin{equation}\label{}
    Q_t \rho = \sum_{i,j=1}^d \pi_{ij} f_j(t) |i\>\<j| \, \rho\, |j\>\<i|\
    ,
\end{equation}
where $f_j(t) =- \dot{n}_j(t)$, that is, $n_j(t)$ may be represented
{\em via}
\begin{equation}\label{}
    n_j(t) = 1 - \int_0^t f_j(\tau) d\tau\ .
\end{equation}
In this case
\begin{equation}\label{}
    b_{ij}(t) = \pi_{ij} \kappa_j(t)\ ,
\end{equation}
where $\kappa_j(t)$ are defined in terms of the Laplace transform
\begin{equation}\label{}
    \widetilde{\kappa}_j(s) =
    \frac{s\widetilde{f}_j(s)}{1-\widetilde{f}_j(s)}\ .
\end{equation}
Inserting into (\ref{NM-CLASS}) one recovers the  old result of
Gillespie \cite{Gillespie} (see also \cite{Klafter} and the
discussion in \cite{B} on continuous-time random walk).


\section{Reducing Schr\"odinger dynamics}  \label{WW}

In this section we provide a simple construction giving rise to the
family of completely positive maps $N_t$ satisfying initial
condition $N_0=\oper$. Consider the unitary evolution in
$\mathcal{H}_S \ot \mathcal{H}_R$ governed by the Hamiltonian $H$.
Let $|\omega\> \in \mathcal{H}_R$ be a fixed vector state of the
reservoir  and let us define the projector
\begin{equation}\label{}
    P \, :\, \mathcal{H}_S \ot \mathcal{H}_R  \longrightarrow \mathcal{H}_S \ot
    \mathcal{H}_R \ ,
\end{equation}
by the following formula
\begin{equation}\label{}
    P = \mathbb{I}_S \ot |\omega\>\<\omega| \,
\end{equation}
that is,
\begin{equation}\label{}
    P (|\psi\> \ot |\phi\>) = |\psi\> \ot \<\omega|\phi\>|\omega\> \ .
\end{equation}
Having defined $P$ one introduces the reduced dynamics in
$\mathcal{H}_S$  by
\begin{equation}\label{}
    |\psi_t\> = n_t |\psi_0\> \ ,
\end{equation}
where the time-dependent evolution operators $n_t : \mathcal{H}_S
\rightarrow \mathcal{H}_S$ is defined by
\begin{equation}\label{}
    n_t \ot |\omega\>\<\omega| = P e^{-itH} P\ ,
\end{equation}
and satisfies $n_0 = \mathbb{I}_S$. In analogy to (\ref{NF}) it can
be represented as follows
\begin{equation}\label{n-nu}
    n_t = \mathbb{I}_S - \int_0^t \nu_\tau d\tau\ ,
\end{equation}
where $\nu_t = - \dot{n}_t$ and hence it satisfies non-local
equation
\begin{equation}\label{n-nmarkov}
    \dot{n}_t = - \int_0^t z_{t-\tau} n_\tau d\tau\ ,
\end{equation}
where the generator is defined in terms of its Laplace transform
\begin{equation}\label{}
    \widetilde{z}_s = \frac{\mathbb{I}_S - s\,\widetilde{n}_s}{\widetilde{n}_s}\ .
\end{equation}
Equivalently, if $|\varphi_t\>$ is a solution of the Schr\"odinger
equation
\begin{equation}\label{}
    i \frac{d |\varphi_t\>}{dt} = H |\varphi_t\> \ ,
\end{equation}
with an initial condition $|\varphi_0\> = |\psi_0\> \ot |\omega\>$,
then $|\psi_t\>$ is nothing but the reduction of $|\varphi_t\>$
\begin{equation}\label{}
    |\psi_t \> \ot |\omega\> = P|\varphi_t\> \ .
\end{equation}
It should be clear from (\ref{n-nmarkov}) that the reduced evolution
$n_t$ does not satisfy Schr\"odinger-like equation in
$\mathcal{H}_S$. Note, that $n_t$ is a contraction in
$\mathcal{H}_S$, that is
\begin{equation}\label{cont}
    \< n_t \psi| n_t \psi\> \leq \<\psi|\psi\> \ ,
\end{equation}
for $t \geq 0$, and hence $n_t$ does not define a legitimate
dynamics of the pure state $|\psi_t\>$ in the system Hilbert space
$\mathcal{H}_S$.  One may easily define normalized solution
\begin{equation}\label{}
    |\psi'_t\> = \frac{|\psi_t\>}{||\psi_t||}\ .
\end{equation}
Note, however, that normalized evolution $|\psi'_t\>$ is, contrary
to $|\psi_t\>$,  no longer linear.

Here, we follow our general approach. Let us define the following
evolution in the space of mixed states
\begin{equation}\label{}
    N_t \rho = n_t \rho\, n_t^\dagger\ .
\end{equation}
By construction $N_t$ is completely positive and satisfies an
initial condition $N_0=\oper$. Interestingly, the inverse
\begin{equation}\label{}
    N_t^{-1} = n_{-t} \rho\, n_{-t}^\dagger\ ,
\end{equation}
does exist for almost all $t \geq 0$ and it is again completely
positive. Hence, if
\begin{equation}\label{}
    - \dot{N}_t^\# \mathbb{I} = n_t^\dagger \nu_t + \nu_t^\dagger
    n_t \geq 0\ ,
\end{equation}
then one can find a family of completely positive maps $Q_t$ and
define the legitimate memory kernel $\mathcal{K}_t = B_t - Z_t$.

\begin{remark} {\em Usually, $H = H_0 + \lambda H_{\rm int}$. It is well known
\cite{Davies-WW} that in the weak coupling limit $|\psi_t\> =
n_t|\psi_0\> $ satisfies  satisfies the following equation
\begin{equation}\label{}
    \dot{n}_t  = -z\, n_t \ ,\ \ \ n_0 = \oper\ ,
\end{equation}
with $z = ih + \frac 12 X$, where $h$ is Hermitian and $X \geq 0$
(hence Hermitian). Clearly, $z$ plays a role of Wigner-Weisskopf
non-Hermitian Hamiltonian. Interestingly weak coupling limit
guaranties that $z$ is a normal operator, i.e. $h$ and $X$ mutually
commute. }
\end{remark}

\begin{Example} {\em Let $n_t$ be defined by its spectral
decomposition
\begin{equation}\label{n-x}
    n_t |k\> = x_k(t)  |k\> \ ,
\end{equation}
and hence one obtains
\begin{equation}\label{}
    N_t \rho = \sum_{k,l} n_{kl}(t) |k\>\<k| \rho |l\>\<l|\ ,
\end{equation}
where
\begin{equation}\label{}
    n_{kl}(t) = {x_k(t)} \overline{ x_l(t)}\ .
\end{equation}
One finds
\begin{equation}\label{}
    N_t^\# \mathbb{I} = \sum_k |x_k(t)|^2 |k\>\<k|\ ,
\end{equation}
which shows that $N_t$ is normalized iff $x_k(t) =
e^{-i\varepsilon_kt}$.
 Consider now
\begin{equation}\label{}
    x_k(t) = e^{-i\varepsilon_k} (1 - \int_0^t f_k(\tau) d\tau)\ .
\end{equation}
Note, that if $f_k(t)$ has a special form
\begin{equation}\label{}
    f_k(t) = \kappa_k e^{-\gamma_k t} \ , \ \ \ \gamma_k \geq 0\ ,
\end{equation}
then
\begin{equation}\label{}
    x_k(t) = \gamma_k^{-1} e^{-i\varepsilon_k} (\gamma_k - \kappa_k + \kappa_k e^{-\gamma_k
    t} ) \ ,
\end{equation}
and hence in the limit $\kappa_k \rightarrow\gamma_k$ one recovers
Wigner-Weisskopf theory
\begin{equation}\label{}
    x_k(t) = e^{-[i\varepsilon_k  + \gamma_k]t}\ .
\end{equation}
}
\end{Example}

\begin{Example} {\em

Consider the pure decoherence model,
\begin{equation}
\label{Htot} H=H_R + H_S + H_{SR},
\end{equation}
where $H_R$ is the reservoir Hamiltonian,
\begin{equation}
\label{HS} H_S = \sum_k \epsilon_k P_k \ ,
\end{equation}
with $P_k = |k\>\<k|$,  the system Hamiltonian and
\begin{equation}\label{}
    H_{SR} = \sum_k P_k \ot B_k
\end{equation}
the interaction part, $B_k=B_k^\dagger$ being reservoirs operators.
The total Hamiltonian has therefore the following form
\begin{equation}\label{}
    H = \sum_k P_k \ot R_k\ ,
\end{equation}
where the reservoir operators $R_k$ read as follows
\begin{equation}\label{}
    R_k = \varepsilon_k \mathbb{I}_R + H_R + B_k \ .
\end{equation}
One easily finds the the reduced dynamics $n_t$ is defined by the
formula (\ref{n-x}) with
\begin{equation}\label{}
    x_k(t) = \< \omega | e^{-i R_k t}| \omega\> = e^{-i\varepsilon_k
    t} \< \omega | e^{-i [H_R + B_k] t}| \omega\>\ .
\end{equation}
The presence of nontrivial factor $\< \omega | e^{-i [H_R + B_k] t}|
\omega\>$ is responsible for all memory effects.

}
\end{Example}

\section{Conclusions}

We have provided a class of legitimate memory kernels leading to
completely positive trace preserving dynamical maps. Our
construction is based on the simple observation that if the family
of completely positive maps $N_t$ with an initial condition
$N_0=\oper$ satisfies an additional condition $\dot{N}^\#_t
\mathbb{I} \leq 0$, then one may perform suitable normalization and
as a result one obtains a family of completely positive trace
preserving maps $\Lambda_t$ generated by the legitimate memory
kernel $\mathcal{K}_t$. This procedure is highly non unique.
Interestingly, when applied to Wigner-Weisskopf theory it gives the
standard Markovian evolution governed by the local master equation.
As a byproduct we have constructed a class of legitimate memory
kernels for classical stochastic non-Markovian dynamics.

\section*{Acknowledgment}

This work was partially supported by the Polish Ministry of Science
and Higher Education Grant No 3004/B/H03/2007/33.

\end{document}